\documentstyle[11pt,epsfig]{article} 
\newcommand{\be}{\begin{equation}}
\newcommand{\ee}{\end{equation}}
\newcommand{\bav}{\begin{eqnarray}}
\newcommand{\eav}{\end{eqnarray}}
\evensidemargin=\oddsidemargin
\title{The Angular Scale of Topologically-Induced Flat Spots 
in the Cosmic Microwave Background Radiation}
\author{David Olson and Glenn D. Starkman \\
Dept. of Physics, Case Western Reserve University, Cleveland, OH 44106}
\begin{document}
\maketitle
\abstract{The notion that the topology of the universe need
not be that of the universal covering space of its geometry has recently received
renewed attention \cite{Starkman}.  Generic signatures of cosmological topology have 
been sought, both in the distribution of objects in the universe,
and especially in the  temperature fluctuations of the cosmic microwave 
background radiation (CMBR).  One signature\cite{LevinA}, identified in the 
horn topology but hypothesized to be generic\cite{LevinA,CSS} is
featureless regions or flat spots  in the CMBR sky.  We show that typical
observation points within the cusped 3-manifold m003 from the Snappea 
census\cite{Snappea} have flat spots with an angular scale of about five 
degrees for $\Omega_0$=0.3.  We expect that this holds for other small 
volume cusped manifolds with this $\Omega_0$ value.}
\vfil

\section{Introduction}

Recently, there has been considerable interest in the prospect that
the universe has non-trivial topology.  This heightened awareness
of the possibility of topology has been driven by the ever strengthening
case against  a flat, matter dominated universe.  
The two alternatives which best fit the data are either a flat,
cosmological constant dominated universe, or a negatively curved,
matter dominated universe.  If the universe has negative curvature,
then the curvature scale is also the natural  scale on which one
would generically expect topology.  

Since the curvature scale of a negatively curved universe
\be
\label{rcurv}
R_c \simeq {3000 h^{-1}\over \sqrt{1-\Omega}}\mathrm{Mpc}
\ee
is considerably less than the radius of the observable universe,
one might hope to be able to observe evidence for the non-trivial 
topology, and comparisons of the COBE satellite's observations 
of fluctuations in the Cosmic Microwave Background 
Radiation (CMBR) temperature to predicted fluctuations for closed 
hyperbolic manifolds have been made \cite{BPS}.
Much effort has gone into uncovering good signatures for 
such topology \cite{Starkman,LR}. The effort is complicated by the fact 
that there are infinitely many different possible topologies for 
negatively curved  manifolds. Moreover, a large, perhaps infinite number
of these have small enough closed loops (in at least some directions)
that they could in principle yield observable consequences.
Therefore the best signatures must allow general searches for 
topology and not just allow us to tell whether the universe
is a particular manifold.

One such generic topology search algorithm utilizes
the very robust observation that the existence of topology  will
result in the existence of pairs of circles on the sky which
have highly correlated patterns of CMBR temperature fluctuations \cite{circles}.

Another suggestion,  made by Levin et al \cite{LevinA}, is that flat spots 
(regions with suppressed long-wavelength fluctuations in temperature)
will appear in the CMBR sky, 
in particular in compact hyperbolic manifolds.
Levin et al examined one particular hyperbolic manifold,
the so-called horn topology (which is not compact),
and discovered that down the direction of the horn
such flat spots do indeed appear.
To understand the suggestion that these flat spots are generic
one  must realize that hyperbolic manifolds of non-trivial topology
correspond to tilings of the   covering space of hyperbolic geometry,
the usual ``open'' $\mathbf{H}^3$.  A compact hyperbolic manifold
is then a tiling whose fundamental or Dirichlet domain
does not extend to spatial infinity.  Such compact hyperbolic
manifolds are however constructed by a process of Dehn surgery
on so-called cusped manifolds, which extend to infinity at
a finite number of isolated points, called cusps.  
The cusp portions of the  cusped manifolds
are very much like the horn topology in that in both cases
the cross-section of the manifold narrows exponentially as one moves 
down the horn/cusp toward spatial  infinity.  As the manifold narrows,
geodesics can readily  wrap around the horn/cusp a large number
of time and so smooth out  any features.
The suggestion  that  the flat  spots seen in the horn topology may be more
generic, assumes that the Dehn surgery required to turn the
cusped manifold into the compact manifold is sufficiently gentle so as to
preserve this evidence of the cusps of the  parent cusped manifold.

In this paper we will try to see just how far the analogy between
cusped and horned manifold can  take us -- how flat are the
flat spots which would appear in the cusped manifold.
This represents  the flattest that one could expect the flat
spots in the daughter cusp-free manifold to be.  We will
show that although the cusps {\em do} produce flat spots
they are generically not quite so prominent as those produced in the horn
topology.

\section{Modes on the Horosphere}

In a small volume cusped hyperbolic manifold, 
it is difficult to calculate the eigenmodes of the wave operator
which contribute to variations  in the CMBR 
(although see \cite{Inoue,Aurich,CornSper,Steil}).  
However, since all that interests us
is whether or not there is a flat spot on the CMBR in the vicinity of a cusp, 
we need not solve the full problem.  
Instead, we find how the topology affects 
the modes on the surface of last scattering (SLS) near the cusp. 

To compute these modes, 
we need to choose a model of $\mathbf{H}^3$ in which to compute.
There are many models of $\mathbf{H}^{3}$.  
The most common representation of $\mathbf{H}^3$ is
Poincare's model, which is  the unit ball in $\mathbf{R}^{3}$ 
with the metric $ds^{2}=\frac{4}{(1-r^{2})^2}dx^{2}$. 
Here $dx^{2}$ is the normal metric of $\mathbf{R}^{3}$ 
and $r$ is the distance from the origin.  
In this model, geodesics are diameters of the unit sphere and 
circular arcs perpendicular to the surface of the unit sphere \cite{Thurston}.  
Of more use to us will be the the hyperboloid model of $\mathbf{H}^3$,
which is the set of points in $\mathbf{R}^{1+3}$ on the upper sheet 
of the hyperboloid $-1=-x_{0}^{2}+x_{1}^{2}+x_{2}^{2}+x_{3}^{2}$.  
The distance $d$ between two points $x,y$ in
this model is $d=$arccosh$(-x\circ y)$, where  
$x\circ y=-x_{0}y_{0} + x_{1}y_{1} + x_{2}y_{2} + x_{3}y_{3}$ 
is the Lorentz dot product of two points in $\mathbf{R}^{1+3}$.  
Geodesics in the hyperboloid model have the
form $\lambda(t)=x\cosh(t)+y\sinh(t)$, where $x$ is a point on the
hyperboloid and $y$ is a unit vector in $\mathbf{R}^{1+3}$ orthogonal to it
\cite{Ratcliffe}.  
Finally, we will also make use of the Klein model of $\mathbf{H}^{3}$.
This is obtained from the hyperboloid model 
by projecting the point $(x_{0},x_{1},x_{2},x_{3})$ to the point
$(\frac{x_{1}}{x_{0}},\frac{x_{2}}{x_{0}},\frac{x_{3}}{x_{0}})$.  
Geodesics in the Klein model are open chords of the unit ball\cite{Thurston}.

We now use a horosphere to find the modes near the cusp.  
A horosphere  is a sphere inside and tangent to the unit sphere in the
Poincare model of $H^3$.  
We consider the horosphere tangent at the cusp that goes
through the point on the SLS in the direction of the cusp.  
On the horosphere, the transformation group of the manifold 
restricts to a Euclidean similarity group\cite{Ratcliffe}.  
We calculate the modes of this group, and use
them as an approximation to the modes on the SLS.  
By comparing the density of these modes 
to those of a corresponding patch of open sky, 
we get an estimate of any suppression caused by the topology. 

\begin{figure}[bt]
  \begin{center}\epsfig{figure=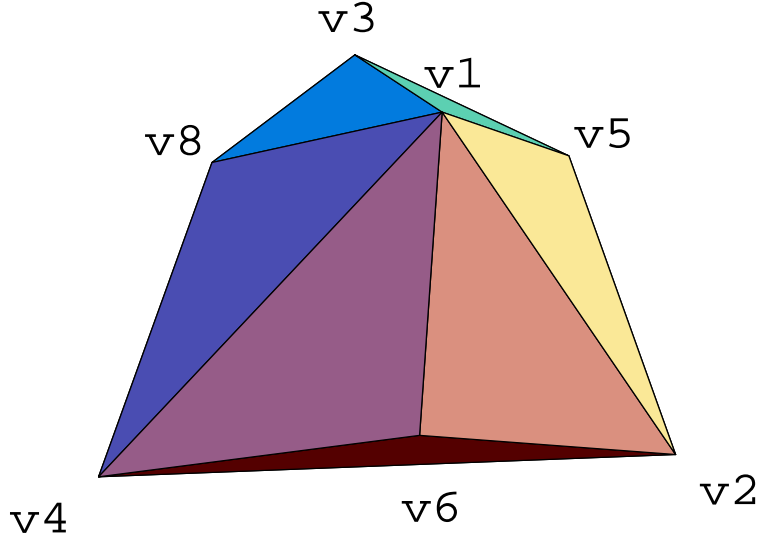,height=3in,width=4in}\end{center}
  \begin{center}\caption{The Dirichlet domain of the manifold m003 in the Snappea
census centered at the origin of the Klein model.}\end{center}
  \label{fig:Dirichlet}
\end{figure}

We did this calculation on a particular cusped manifold  --
number m003 from the Snappea census of cusped manifolds\cite{Snappea}.
This manifold is obtained by gluing the faces of two ideal tetrahedra%
\footnote{An {\it ideal} polyhedron is one with only ideal vertices.
An ideal vertex in the Poincare or Klein model
is a vertex located on the unit sphere. A {\it finite} vertex is a vertex
which is not ideal.}
together and has a volume $V\approx 2.0299$, in units of the curvature 
radius cubed.  The Dirichlet domain we considered (cf. figure \ref{fig:Dirichlet})
is centered on one of the tetrahedra, 
with the other tetrahedron split into quarters which
are attached to the faces of the first tetrahedron.  The resulting 
figure has four ideal vertices, four finite vertices, eighteen edges, 
and twelve faces.
Numbering the ideal vertices 1, 2, 3, and 4, we can associate each
finite vertex with the ideal vertices which it shares edges with.
The finite vertices are numbered 5, 6, 7, and 8, 
with vertex 5 forming edges with vertices 1, 2, and 3, 
vertex 6 forming edges with vertices 1, 2, and 4,
vertex 7 forming edges with vertices 2, 3, and 4,
and vertex 8 forming edges with vertices 1, 3, and 4.
Each face can be identified by its three vertices.  
In m003, the faces are glued in the pattern 
125-347, 237-348, 138-246, 148-135, 146-247, and 126-235.  
There are then six classes of faces, 
six classes of edges and two classes of vertices.

In the Klein model, the vertices of m003 are
\bav
v_{1}&=\left(-\frac{1}{\sqrt{3}},-\frac{1}{\sqrt{3}},\frac{1}{\sqrt{3}}
\right),
\quad v_{2}=\left(\frac{1}{\sqrt{3}},-\frac{1}{\sqrt{3}},-\frac{1}{\sqrt{
3}}\right), 
\nonumber\\
v_{3}&=\left(\frac{1}{\sqrt{3}},\frac{1}{\sqrt{3}},\frac{1}{\sqrt{3}} \right),
\quad 
v_{4}=\left(-\frac{1}{\sqrt{3}},\frac{1}{\sqrt{3}},-\frac{1}{\sqrt{3} }\right),
\\v_{5}&=\left(\frac{\sqrt{3}}{5},-\frac{\sqrt{3}}{5},\frac{\sqrt{3}}{5}
\right),
\quad v_{6}=\left(-\frac{\sqrt{3}}{5},-\frac{\sqrt{3}}{5},-\frac{\sqrt{3}}{5}
\right),
\nonumber\\
v_{7}&=\left(\frac{\sqrt{3}}{5},\frac{\sqrt{3}}{5},-\frac{\sqrt{3}}{5}
\right),
\quad {\mathrm{and}}\quad 
v_{8}=\left(-\frac{\sqrt{3}}{5},\frac{\sqrt{3}}{5},\frac{\sqrt{3}}{5}
\right).
\nonumber
\eav
The six generators of the transformation group are, in the
 hyperboloid model,
\bav
a_{0}=\left(\begin{array}{cccc} 
\!\frac{7}{4} &\! -\frac{\sqrt{3}}{4} &\! -\frac{3\sqrt{3}}{4} 
	&\! \frac{\sqrt{3}}{4} \\
\!\frac{\sqrt{3}}{4} &\! -\frac{1}{4} &\! -\frac{3}{4} &\!-\frac{3}{4} \\
\!-\frac{3\sqrt{3}}{4} &\! \frac{3}{4} &\! \frac{5}{4} &\! -\frac{3}{4} \\
\!\frac{\sqrt{3}}{4} &\! \frac{3}{4} &\! -\frac{3}{4} &\! \frac{1}{4}
 \end{array}\!\right),\quad
a_{1}=\left(\begin{array}{cccc} 
\!\frac{7}{4} &\! \frac{\sqrt{3}}{4} &\! -\frac{3\sqrt{3}}{4} &
\!-\frac{\sqrt{3}}{4} \!\\
\!\frac{3\sqrt{3}}{4} &\! \frac{3}{4} &\! -\frac{5}{4} &\!-\frac{3}{4} \!\\
\!\frac{\sqrt{3}}{4} &\! -\frac{3}{4} &\! -\frac{3}{4} &\! -\frac{1}{4} \!\\
\!-\frac{\sqrt{3}}{4} &\! -\frac{1}{4} &\! \frac{3}{4} &\! -\frac{3}{4}\!
 \end{array}\!\right),
\nonumber\\
a_{2}=\left(\begin{array}{cccc} 
\!\frac{7}{4} &\! \frac{\sqrt{3}}{4} &\! \frac{\sqrt{3}}{4} 
	&\! \frac{3\sqrt{3}}{4} \\
\!-\frac{\sqrt{3}}{4} &\! -\frac{1}{4} &\! \frac{3}{4} &\!-\frac{3}{4} \\
\!\frac{\sqrt{3}}{4} &\! -\frac{3}{4} &\! \frac{1}{4} &\! \frac{3}{4} \\
\!\frac{3\sqrt{3}}{4} &\! \frac{3}{4} &\! \frac{3}{4} &\! \frac{5}{4}
 \end{array}\!\right),\quad
a_{3}=\left(\begin{array}{cccc} 
\!\frac{7}{4} &\! -\frac{\sqrt{3}}{4} &\! \frac{\sqrt{3}}{4} &
\!-\frac{3\sqrt{3}}{4} \!\\
\!-\frac{3\sqrt{3}}{4} &\! \frac{3}{4} &\! -\frac{3}{4} &\!\frac{5}{4} \!\\
\!\frac{\sqrt{3}}{4} &\! -\frac{1}{4} &\! -\frac{3}{4} &\! -\frac{3}{4} \!\\
\!\frac{\sqrt{3}}{4} &\! \frac{3}{4} &\! \frac{1}{4} &\! -\frac{3}{4}\!
 \end{array}\!\right),
\\
a_{4}=\left(\begin{array}{cccc} 
\!\frac{7}{4} &\! -\frac{\sqrt{3}}{4} &\! -\frac{\sqrt{3}}{4} 
	&\! \frac{3\sqrt{3}}{4} \\
\!-\frac{3\sqrt{3}}{4} &\! \frac{3}{4} &\! \frac{3}{4} &\!-\frac{5}{4} \\
\!-\frac{\sqrt{3}}{4} &\! \frac{1}{4} &\! -\frac{3}{4} &\! -\frac{3}{4} \\
\!-\frac{\sqrt{3}}{4} &\! -\frac{3}{4} &\! \frac{1}{4} &\! -\frac{3}{4}
 \end{array}\!\right)\quad {\mathrm{and}} \quad
a_{5}=\left(\begin{array}{cccc} 
\!\frac{7}{4} &\! -\frac{3\sqrt{3}}{4} &\! \frac{\sqrt{3}}{4} &
\!-\frac{\sqrt{3}}{4} \!\\
\!-\frac{\sqrt{3}}{4} &\! \frac{3}{4} &\! \frac{3}{4} &\!\frac{1}{4} \!\\
\!-\frac{3\sqrt{3}}{4} &\! \frac{5}{4} &\! -\frac{3}{4} &\! \frac{3}{4} \!\\
\!-\frac{\sqrt{3}}{4} &\! \frac{3}{4} &\! -\frac{1}{4} &\! -\frac{3}{4}\!
 \end{array}\!\right).\nonumber
\eav
These generators are orientation preserving and satisfy the six relations
\bav
a_{0}a_{1}^{-1}a_{5}^{-1}=1,\quad a_{3}a_{2}a_{4}^{-1}=1,
\nonumber\\
a_{1}a_{2}a_{4}=1,\quad a_{0}a_{1}a_{3}=1,
\\
a_{5}a_{0}a_{4}^{-1}=1,\quad {\mathrm{and}} \quad a_{2}a_{5}a_{3}^{-1}=1. 
\nonumber
\eav

\begin{figure}[bt]
  \begin{center}\epsfig{figure=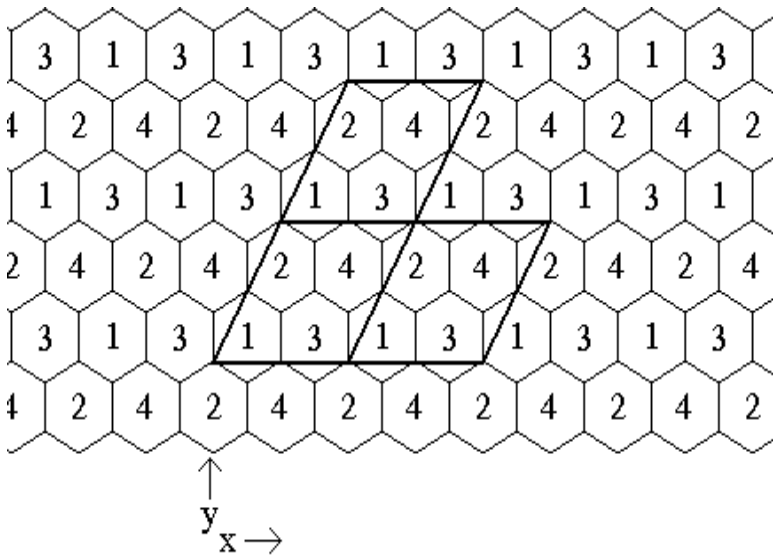,height=3in,width=4in}\end{center}
  \begin{center}\caption{Hexagonal tiling of the horosphere with the rectangular
 domain used to calculate the modes of the wave equation superimposed.}\end{center}
  \label{fig:triangles}
\end{figure}

The Euclidean similarity group of a horosphere centered on a cusp 
is a tiling of the Euclidean plane with hexagons.  
There are four elements in the tiling, 
one from each ideal vertex of the domain.  
The tiling is shown in figure 1.  
To calculate the eigenmodes of the wave operator on the tiling, 
the rectangular domain and the axes shown in figure 1 were used.  
The similarity group of this tiling is generated by the two transformations
$T_{1}(x,y)=(x+L,y)$, $T_{2}(x,y)=(x+\frac{L}{2},y+\frac{\sqrt{3}L}{2})$, and  
the modes $\phi(x,y)$ of the tiling are
solutions of the Helmholtz (wave) equation under the two boundary conditions 
$\phi(T_i(x,y))=\phi(x,y)$ for $i=1,2$.
A straightforward argument shows that the normal modes that satisfy these
boundary conditions have wavevectors $\vec{k}$ of the form 
\be 
\vec{k}=\frac{2\pi}{L}\left[n\left(1,-\frac{1}{\sqrt{3}}\right)
+m\left(0,\frac{2}{\sqrt{3}}\right)\right]
\ee
with $n$, $m$ arbitrary integers.  The shortest nonzero $\vec{k}$ modes
have $k_{min}=\frac{2\pi}{L}\frac{2}{\sqrt{3}}$.
This corresponds to a maximum wavelength 
$\lambda_{max}=\frac{2\pi}{k_{min}} = \frac{\sqrt{3}L}{2}$ for solutions 
with these boundary conditions.

\section{From Horosphere to Sphere of Last Scatter}
We have now found the wavelengths of the modes
on the portion of the SLS that is tiled like the horosphere.  
To find $L$ and calculate how much of the SLS is tiled, 
we computed the intersections of the SLS with 
the edges of the domains in the tiling.  
The size and maximum wavelength of flat spots on the CMB varies 
from point to point in the manifold, depending on how
far down the cusp a point is.  To account for this,
we chose to consider the SLS of three points: 
the point $(1,0,0,0)$ in the hyperboloid model, 
the point at the radius of half volume ($0.61$ in units of the curvature
radius $R_{curv}$) towards the cusp from 
$(1,0,0,0)$, 
and the point at a distance equal to the curvature radius
towards the cusp from $(1,0,0,0)$.
The point $(1,0,0,0)$ is at the center of the Dirichlet domain we considered,
at a distance $0.58 R_{curv}$ from the nearest face.  It is positioned 
as far from the small regions down the cusp as possible, so it 
has the smallest amount of its SLS tiled and the largest cusp wavelength 
cutoff of the points in the manifold. 
The radius of half volume is calculated by determining the radius of 
a sphere centered on $(1,0,0,0)$ whose intersection with the
Dirichlet domain has a volume half that of the manifold.
The point at the radius of half volume is closer to the small cusp regions
than approximately half of the points in the manifold, 
so it has a spot size and wavelength cutoff which we take to be 
representative of an average point of the manifold.  It is a distance 
$0.33 R_{curv}$ from the nearest face.
The point at the curvature radius distance 
is well down the cusp, with a distance of only $0.22 R_{curv}$ from the 
nearest face, so we will use it to estimate the spot size and shortest 
wavelength cutoff of points in the manifold which are further down the
cusp than average.
The radius of the last scattering surface 
in units of the curvature scale\cite{circles} is a function of $\Omega_0$:  
\be
R_{SLS}\approx R_{curv}\mathrm{arccosh}\left(\frac{2-\Omega _{0}}{\Omega _{0}}
\right) .
\ee
Using $\Omega _{0}=0.3$ and $R_{curv}\equiv 1$, 
we get $R_{SLS}\approx 2.4$. 
(This is actually the radius of the particle horizon, which is marginally larger
than the radius of the SLS.)

In the domain centered on $(1,0,0,0)$ in the hyperboloid model, a point
 $p$ on the edge between the ideal vertices $v_{i}$ and $v_{j}$ 
 satisfies the equation $p(r)=e_{ij}\cosh r + t_{ij} \sinh r$, 
 where $e_{ij}$ is the center of the edge $ij$ and $t_{ij}$ is the unit 
 vector tangent to the hyperboloid pointing along the edge.  
The $e_{ij}$ and $t_{ij}$ are:
\be
\begin{array}{ll}
 e_{12}=\left(\sqrt{\frac{3}{2}},0,-\sqrt{\frac{1}{2}},0\right),\quad  
&t_{12}=\left(0,-\sqrt{\frac{1}{2}},0,\sqrt{\frac{1}{2}}\right)\\
 e_{13}=\left(\sqrt{\frac{3}{2}},0,0,\frac{1}{\sqrt{2}}\right)\quad
&t_{13}=\left(0,-\frac{1}{\sqrt{2}},-\frac{1}{\sqrt{2}},0\right)\\
 e_{14}=\left(\sqrt{\frac{3}{2}},-\frac{1}{\sqrt{2}},0,0\right)\quad
&t_{14}=\left(0,0,-\frac{1}{\sqrt{2}},\frac{1}{\sqrt{2}}\right) \\
 e_{23}=\left(\sqrt{\frac{3}{2}},\frac{1}{\sqrt{2}},0,0\right),\quad
&t_{23}=\left(0,0,-\frac{1}{\sqrt{2}},-\frac{1}{\sqrt{2}}\right)\\
 e_{24}=\left(\sqrt{\frac{3}{2}},0,0,-\frac{1}{\sqrt{2}}\right)\quad
&t_{24}=\left(0,\frac{1}{\sqrt{2}},-\frac{1}{\sqrt{2}},0\right)\\
 e_{34}=\left(\sqrt{\frac{3}{2}},0,\frac{1}{\sqrt{2}},0\right)\quad 
&t_{34}=\left(0,\frac{1}{\sqrt{2}},0,\frac{1}{\sqrt{2}}\right)
\end{array}
\ee

The edges between a finite vertex $v_{i}$ and an ideal vertex $v_{j}$
satisfy the equation $p(r)=v_{i}\cosh r + t_{ij} \sinh r$, where
$v_{i}$ is the finite vertex and $t_{ij}$ is the unit vector tangent
to the hyperboloid in the direction of the ideal vertex $v_{j}$.
The $v_{i}$ are stated above in equation 2.  The $t_{ij}$ are:
\be
 \begin{array}{ll}
t_{51}=\left(-\frac{1}{4},-\frac{7}{4\sqrt{3}},-\frac{1}{4\sqrt{3}},
    \frac{1}{4\sqrt{3}}\right) ,\quad &
 t_{52}=\left(-\frac{1}{4},\frac{1}{4\sqrt{3}},-\frac{1}{4\sqrt{3}},
    -\frac{7}{4\sqrt{3}}\right) \nonumber \\
 t_{53}=\left(-\frac{1}{4},\frac{1}{4\sqrt{3}},\frac{7}{4\sqrt{3}},
    \frac{1}{4\sqrt{3}}\right),\quad &
 t_{61}=\left(-\frac{1}{4},-\frac{1}{4\sqrt{3}},-\frac{1}{4\sqrt{3}},
    \frac{7}{4\sqrt{3}}\right) \nonumber \\
 t_{62}=\left(-\frac{1}{4},\frac{7}{4\sqrt{3}},-\frac{1}{4\sqrt{3}},
    -\frac{1}{4\sqrt{3}}\right),\quad &
 t_{64}=\left(-\frac{1}{4},-\frac{1}{4\sqrt{3}},\frac{7}{4\sqrt{3}},
    -\frac{1}{4\sqrt{3}}\right) \\
 t_{72}=\left(-\frac{1}{4},\frac{1}{4\sqrt{3}},-\frac{7}{4\sqrt{3}},
    -\frac{1}{4\sqrt{3}}\right),\quad &
 t_{73}=\left(-\frac{1}{4},\frac{1}{4\sqrt{3}},\frac{1}{4\sqrt{3}},
    \frac{7}{4\sqrt{3}}\right) \nonumber \\
 t_{74}=\left(-\frac{1}{4},-\frac{7}{4\sqrt{3}},\frac{1}{4\sqrt{3}},
    -\frac{1}{4\sqrt{3}}\right),\quad &
 t_{81}=\left(-\frac{1}{4},-\frac{1}{4\sqrt{3}},-\frac{7}{4\sqrt{3}},
    \frac{1}{4\sqrt{3}}\right) \nonumber \\
 t_{83}=\left(-\frac{1}{4},\frac{7}{4\sqrt{3}},\frac{1}{4\sqrt{3}},
    \frac{1}{4\sqrt{3}}\right),\quad &
 t_{84}=\left(-\frac{1}{4},-\frac{1}{4\sqrt{3}},\frac{1}{4\sqrt{3}},
    -\frac{7}{4\sqrt{3}}\right) \nonumber 
\end{array}
\ee

Every domain in the tiling can be obtained by applying an element of
 the transformation group of the tiling to the domain centered at
 $(1,0,0,0)$. Applying the transformation of a domain to these edge
 parametrizations yields a parametrization of the edges of that domain.

\begin{figure}
  \begin{center}\epsfig{figure=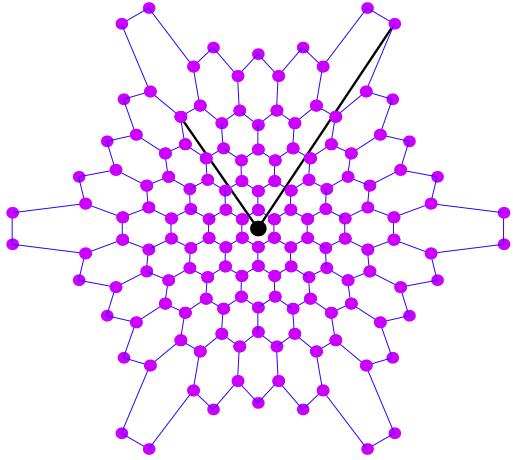,height=3in}\end{center}
\vspace{-1in}
  \caption{Tiling of space as projected onto the night sky.}
  \label{fig:night}
\end{figure}
The intersection of any edge with the SLS of any point $P_{c}$ 
 can now be found by numerically
 solving the equation $D(p(r),P_{c})=2.4$, where $D(x,y)$ is the
 distance between two points in the hyperboloid model and $p(r)$ is the
 parametrization of the edge.  All intersections of edges with the SLS of 
 each of the three points in
 the vicinity of cusp 1 of the domain centered at $(1,0,0,0)$ were
 calculated.  The results for the point $(1,0,0,0)$ are shown in figure 2,
 which plots the pattern of
 domain edge intersections as seen on the night sky.  The connected points are
 intersections of the edges of the Dirichlet domain with the SLS.  
 The large dot in the center is the geodesic traveling straight down the cusp.  
 The line segments from the cusp show the scale of the diagram, 
 and are of length $4.8$ and $8.8$ degrees.  The diagram shows that 
 the tiling of the sky within a half-angle of $4.8$ degrees has little 
 distortion and will have modes similar to the horosphere.  The outer
 hexagons shown are falling back into the large part of the manifold,
 and can no longer be approximated by tiled horosphere hexagons.  The side
 length of the central hexagon is $0.65$ degrees.  The half volume 
 point and the curvature radius point have disks with radii of $5.4$ and 
 $6.2$ degrees and central hexagons with side lengths of $0.35$ and
 $0.24$ degrees.  

We used these hexagon side lengths $l_{h}$ for the scale $L=2\sqrt{3}l_{h}$ 
 of the tiling of the horosphere
 to find $k_{min}=\frac{2\pi}{L}\frac{2}{\sqrt{3}}$ and 
 $\lambda _{max}=\frac{2\pi}{k_{min}}$ for the modes on the horosphere.  The
 point $(1,0,0,0)$ has a $\lambda _{max}=2.0$ degrees, so the longest 
 wavelength mode on the SLS for the point $(1,0,0,0)$ in the vicinity
 of the cusp is approximately $2.0$ degrees.  

In the absence of non-trivial topology, 
 this region will have the modes of a disc of
 radius $4.8$ degrees, which have wavelengths $\lambda _{n}=\frac{2 \pi
 4.8}{J_{m,n}}$ degrees, where $J_{m,n}$ is the $n$th zero of the $m$th
 cylindrical Bessel function.  The longest
 wavelength mode will have $\lambda_{max}=12.5$ degrees. 

Comparing the longest wavelength of the drum, $12.5$ degrees, and
 the longest wavelength of the horosphere, $2.0$ degrees, shows
 that the topology reduces the longest wavelength to about $0.16$ of its 
 expected value.  So the SLS of the point $(1,0,0,0)$
 exhibits a flat spot of approximately $4.8$ degrees. 

The half volume point has a longest horosphere wavelength of $1.05$  
 degrees and a longest disc wavelength of $14.1$ degrees.
 Here the longest wavelength is reduced to about $0.07$ of its expected 
 value on a spot of half angle $5.4$ degrees.  

The curvature radius point has horosphere modes of longest wavelength 
 $0.72$ degrees and disc modes of longest wavelength $16.1$ degrees.  
 The longest wavelengths is reduced to $0.04$ of its normal value on
 a spot of half angle $6.2$ degrees.  
 
These results depend on having a relatively small value of $\Omega_{0}$.  
As an example of this, the corresponding calculations using this method
with an $\Omega_{0}=0.9$, which has an $R_{sls}\approx 0.65 R_{curv}$ 
yield a null result for flattening.  The point 
$(1,0,0,0)$ has only one hexagon in its tiling, with a sidelength of $40$
degrees.  The half volume radius point has seven hexagons in its tiling,
with a central side length of $19$ degrees and serious distortions of the 
outer hexagons.  
The curvature radius point also has a distorted tiling of seven
hexagons, with a central side length of $11.4$ degrees.  
None of these points exhibits an extensive regular tiling of the SLS in 
the direction of the cusp, so our calculations do not predict a flat spot due 
to the cusp for $\Omega_{0}=0.9$.

Finally,  Gaussian random fields do have flat spots, arising
purely from  statistical fluctuations.
However, in order for a statistical flat spot to be confused 
with one of topological origin, 
many modes would have to have an amplitude much smaller than the mean.
For example, in order to create a spot in which the wavelength of the
longest observed mode is only 7\% of the expected value
approximately $600$ modes would have to have statistically small amplitudes.
(Since $\pi\times\left(\frac{100}{7}\right)^2\simeq600$.)

\section{Conclusion}

In reference \cite{LevinA}, the horn topology was shown to have flat spots which could,
 in principle cover a large portion of the CMBR.  By estimating the modes on
 the surface of last scatter for a point at half volume down the cusp, we have 
 found that the cusped manifold m003 from the Snappea census has a flat spot of 
 about five degrees with longest wavelengths cut to about $0.07$ of normal 
 when $\Omega_0=0.3$ for an average observer.  Calculations with two other 
 points show that the flat spot is larger and the wavelength cutoff is more 
 pronounced at points farther down the cusp and smaller and less pronounced for 
 points nearer the center of the manifold.  The calculations suggest that
 similar spots will be seen in any cusped manifold at points which are close
 enough to a cusp.  A cusped manifold will only be able to avoid having spots
 by being large enough that most points are far from cusps, so any small volume
 cusped hyperbolic manifold should have observable flat spots. Such flat spots
are unlikely to be mere statistical fluctuations of the temperature field.
This supports visible flat spots in the CMBR fluctuation maps as a likely, though not
 necessarily automatic, feature in a hyperbolic universe with non-trivial
 topology.

\end{document}